\documentclass[conference]{IEEEtran}
\IEEEoverridecommandlockouts
\usepackage{cite}
\usepackage{amsmath,amssymb,amsfonts}
\usepackage{algorithmic}
\usepackage{graphicx}
\usepackage{textcomp}
\usepackage{xcolor}
\usepackage{float}

\def\BibTeX{{\rm B\kern-.05em{\sc i\kern-.025em b}\kern-.08em
    T\kern-.1667em\lower.7ex\hbox{E}\kern-.125emX}}
\begin{document}

\title{Design of CANSAT for Air Quality Monitoring at an Altitude of 900m \\
}

\author{\IEEEauthorblockN{1\textsuperscript{st} Soma Kunal Raj}
\IEEEauthorblockA{\textit{Electronics and Communication Engineering} \\
\textit{Vidya Jyothi Institute of Technology}\\
Hyderabad,India \\
kunalraj2003@gmail.com}
\and
\IEEEauthorblockN{2\textsuperscript{nd} Yalamanchili Surya Teja}
\IEEEauthorblockA{\textit{Computer Science Engineering} \\
\textit{Vidya Jyothi Institute of Technology}\\
Hyderabad,India \\
suryatejaprogramming@gmail.com}
}

\maketitle

\begin{abstract}
A CANSAT is a can-shaped satellite that is a replica of the mega satellites. The air quality monitoring plays a crucial role in this present scenario because of the growing pollution. The proposed CANSAT named NAMBI-VJ is a cylindrical structure that has a dimension of 310mm in height and 125mm in Diameter. The primary objective of CANSAT is to design a satellite that weighs 1KG and monitor the air quality of the earth’s surface. The secondary objective is to design a mechanical gyroscope to stabilize the CANSAT. A spill-hole parachute is used for the descent control with a rate of 3m/s.  Experiments were conducted by dropping the satellite from an altitude of 900m through a Drone. It was found that the vital parameters like PPM, longitude,CO2 latitude,etc were received from the satellite through Zigbee Communication on a ground station.

\end{abstract}


\section{Introduction}
CANSAT is a miniaturization of a real satellite that performs all  the operations of a real satellite. Students can examine the difficulties encountered in the real satellite by designing a CANSAT. The satellite uses integrated components for collecting data from the atmosphere. The most crucial part of creating the CANSAT is choosing components that fit within the weight and size restrictions. It also helps in analyzing the reasons of the success or failure of the mission, the results of which can be inferred to build a new CANSAT with a different mission. Over the years many attempts were made in the realization of CANSAT.

The CANSAT includes all the parts found in satellites. These CANSATs are used to monitor the layer of air around the Earth. These are placed at lower heights and the tallest they can go is 1 kilometer. These CANSATs use various communication modules to send and receive information. The space satellites use different frequency bands such as C-band, L-band, X-band, and more. These CANSATs use VHF/UHF frequencies such as 433 MHz/863 MHz for communication. They also use RF which has a good communication range. The Lora has a range of 2Km but is not very accurate. In Zigbee communication, the frequency used is between 915.6 and 927.6 MHz, which falls under the UHF category and is used in private mode. 

The CANSAT is useful for learning about space satellites, which can contribute to the development of the space industry. These CANSATs are created by student groups to help people learn about satellites in space. The CANSAT is a small cylindrical model designed for easy deployment in drones and rocket models. They are affordable and simple to build and operate compared to other types of satellites. 

The CANSATs are a simple way to get started in the Space industry. These tools will improve how satellites are launched and used from rockets, making it easier to understand and deploy them in space.

\section{Literature Survey}

This paper \cite{8985514} gives outlines of the structure and payload for CANSAT. The main mission of the design payload is to provide a safe landing carrying all the components safely without any damage. The structure was made from PLA+ carbon fiber that makes it light and strong. The descent control for a container is a parachute which opens after the deployment phase. The auto gyro ensures that the CANSAT lands safely on the ground once the necessary altitude is reached. Sensors are used to measure the environmental parameters. The data from sensors will be collected and transmitted to ground station via XBEE-PRO-S1. The telemetry data is displayed on the GUI. \cite{10074552}This project involves designing a miniature satellite that demonstrates essential satellite functions like telemetry, receiving commands and data collection. The CANSAT is released at a height of 725m which consists of two payloads. The container uses a parachute for descent, while the payloads use a maple seed mechanism. The payloads function as auto rotating payloads and are made to resemble the aerodynamics of a maple seed. XBEE radios are used to send the sensor data to the ground station. The data obtained from the CANSAT is formatted and shown on that computer via a GUI program. \cite{6581344} The main objective to design this satellite named Vecihi was to safely deploy and land payload and to protect the egg inside the payload. They used a quad-copter model for active descent control. They used carbon fibre for more durability and shock resistance. Ardupilot Mega APM2 is used as controller to optimize system. The systems communication component consists of an XBEE Pro S2B model. \cite{10212104}The study focuses on providing a comparative study between Arduino and Vega processors TheJas32 when used with various sensors like pressure, gas, PIR, and temperature, which are having some applications related to space technology. It discovers that Vega has compatibility problems at higher baud rates, while Arduino operates better at these rates. The report highlights the protocols of serial communication, such as 12C, UART, and SPI. \cite{8399591} The structure of the CANSAT is made using 3D print technology. All electronic components are covered by two ABS elements and the rocket SkySnake 82 is used for carrying the CANSAT. To lift the payload Wessex Rocket motor is used. The solid fuel is used in rockets. And for the deployment they employed parachute. \cite{10528162} The main goal of AathreyaSat is to meet the guidelines set by WCRC. The primary mission of this CANSAT is to insert the satellite into the sounding rocket and it should be deployed at an altitude of 500m and it is tasked to measure the air pollution. They designed the parachute of ripstop nylon and employed triple circular parachute for safe landing. At ground station LoRa Ra-02 receiver module is used to collect the data. \cite{8994718} The aim of the designed CANSAT was to take the atmospheric data and images. Its primary mission is to sense atmospheric parameters like temperature, pressure, and altitude, and send real-time data to a ground station. The secondary mission involves capturing and storing images. The CANSAT features a parachute-based recovery system that allows recovery at any height and time. The data was communicated successfully and was displayed on the GUI in ground station system. The camera captured the images during the entire time of flight.\cite{8002947}The design and validation of a novel descent control system with a Mars glider concept was presented in this presentation for the 2016 International CANSAT Competition. This study aims to design a descent control system for a CANSAT that simulates a sensor payload travelling through a Martian atmosphere and sampling the atmospheric composition during flight. The descent control system involves two parts, parachute and glider system the glider unfolding at 400 meters altitude to collect and send telemetry data.\cite{10245113} The VJIT CANSAT is a trash can sized satellite designed for air quality monitoring   at an altitude of 300m. The structure is made of poly lactic acid material developed by 3D printer.  They used spill hole parachute for deployment mechanism because of its high drag coefficient and durability. The telemetry data from the CANSAT was collected at the Ground Station with the help of LoRa communication module.\cite{7362931}The goal of the CANSAT project is to create a nanosatellite for the purpose of gathering telemetry data on air pollutants in Veracruz-Boca del Rio, specifically carbon monoxide, methane, benzene, and butane. The system includes an Arduino Nano, various sensors and an RF transmitter. It features a parachute for controlled descent. The ground station is equipped with an RF receiving antenna and a computer interface for real-time data processing and storage. \cite{10441654}The paper outlines the design and development of a CANSAT for environmental monitoring and scientific exploration, focusing on sensors, telemetry, and descent systems. The telemetry data is also transmitted in real-time to the ground station using the MQTT protocol. An unique aerodynamic rotor mechanism with a 3 axis mechanical gyroscopic system is placed in the CANSAT to control the stabilisation of the satellite. A two-stage parachute system used for safe descent. \cite{10436175} The main objective is to design a decent control system. It used parachute for initial descent and glider for controlled descent considering shock free and safe landing. They also used an autogyro mechanism for orientation of the system. The electrical architecture in this system is divided into two, primary load and secondary load. Primary where egg is being protected and secondary load where it includes microcontroller, sensors, battery  and  actuator. \cite{9231031} The main purpose of this project is to design and implement an affordable CanSat for Bangladeshi students. The deployment of CANSAT is done using the parachute. The cylindrical body was designed using PVC pipes with suitable caps fitted in both the terminals which tolerates shocks generated by collision between CANSAT and the ground. They used a GUI interface in order to collect the data and finally the CANSAT was designed within 50 dollars . \cite{10398763}The project aims to develop a low-cost, open-source communication architecture for scientific CANSATs using Commercial-Off-The-Shelf (COTS) components. It employs Amplitude Shift Keying (ASK) and Pseudo Morse Code (PMC) for data transmission, making it accessible to amateur radio operators. CANSATs are equipped with An Arduino Nano and additional sensors are installed on CANSAT to gather environmental data.\cite{10374870}The paper presents a new modular design philosophy for Scientific CANSATs, calling for fast integration of different scientific instruments through COTS products and open-source technologies. This means a single power and communication bus, mechanical attachments, and distributed computing and data management. The modular CANSATs framework was tested against a set of experiments that showed fast assembly and reconfiguration, thus improving the possibility of research into near-space technology.

\section{Architecture of the CANSAT}
The Cansat consists of the:
\begin{enumerate}
    \item Payload
    \item Structural Subsystem
    \item On-Board Communication
    \item On-Board Computer
    \item Descent Control Subsystem
    \item Power Subsystem
    \item Ground Station
\end{enumerate}
\subsection{Payload Subsystem}\label{A}

The main objective of the payload system is to have a gyroscope stabilisation mechanism which is used to make the cansat have a soft landing.
The secondary objective is to measure the concentration of Gases like CO2 in parts per million (ppm) over the vicinity of 900 meters approximately using an MQ135 sensor. Also, the location of data acquisition is done by GPS Neo6m.
\subsection{Structural Subsystem}
The structural subsystem aims to provide a rigid structure that sustains the flight at an altitude of 900m from Ground Level. The Structure of the NAMBI-VJ CANSAT is made of Poly Lactic acid material developed by the 3D printer facility of the college available in-house. The Poly Lactic Acid is lightweight compared to other materials and hence chosen for the body of the satellite.The dimensions of the NAMBI-VJ CANSAT have a length of 310mm with a diameter of 125mm and a thickness of 3mm. The CAD model of the Satellite is shown in figure 2. The Volume of the CAN is 3,804.27 cm3. The satellite weighs 727.6 g as indicated in table 1
\begin{figure}
\includegraphics[width=1\linewidth]{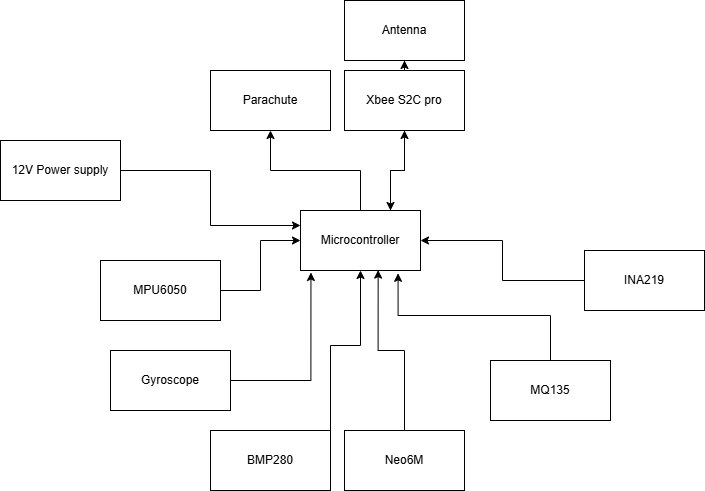}
\caption{Block Diagram of can-satellite}
\label{fig:enter-label}
\end{figure}

\begin{table}[h!]
    \centering
    \begin{tabular}{|c|c|c|}
    \hline
    S.no&Name of the component&Mass \\
    \hline
    1 & Can Structure &300g \\
    2 & Xbee S2C Pro &15g    \\
    3 & MPU-6050 &5g  \\
    4     &MQ135  & 10g \\
    5    &BME180  & 3g \\
    6   &INA219  & 6g \\
    7    &SD card module  &15g  \\
    8    &Neo-6M  &10g  \\
    9    &Arduino Uno  & 25g \\
    10    &Servo Motors x2  & 40g \\
    11  &Buzzer &5.3g  \\
    12  & Pla Sheet&  25.3g \\
    13  &Arduino nano &15g   \\
    14 &BO motors x3 & 100g \\
    15 &Camera Module &40g  \\
    16 &Voltage regulator module & 10g \\
    17 &Battery &100g  \\
    18 &Switch & 3g \\
    &TOTAL & 727.6g \\
    \hline
    
    \end{tabular}
    \newline

    \caption{List of the major components in CANSAT}
   
    \label{tab:my_label}
\end{table}
\begin{figure}
    \centering
    \includegraphics[width=1\linewidth]{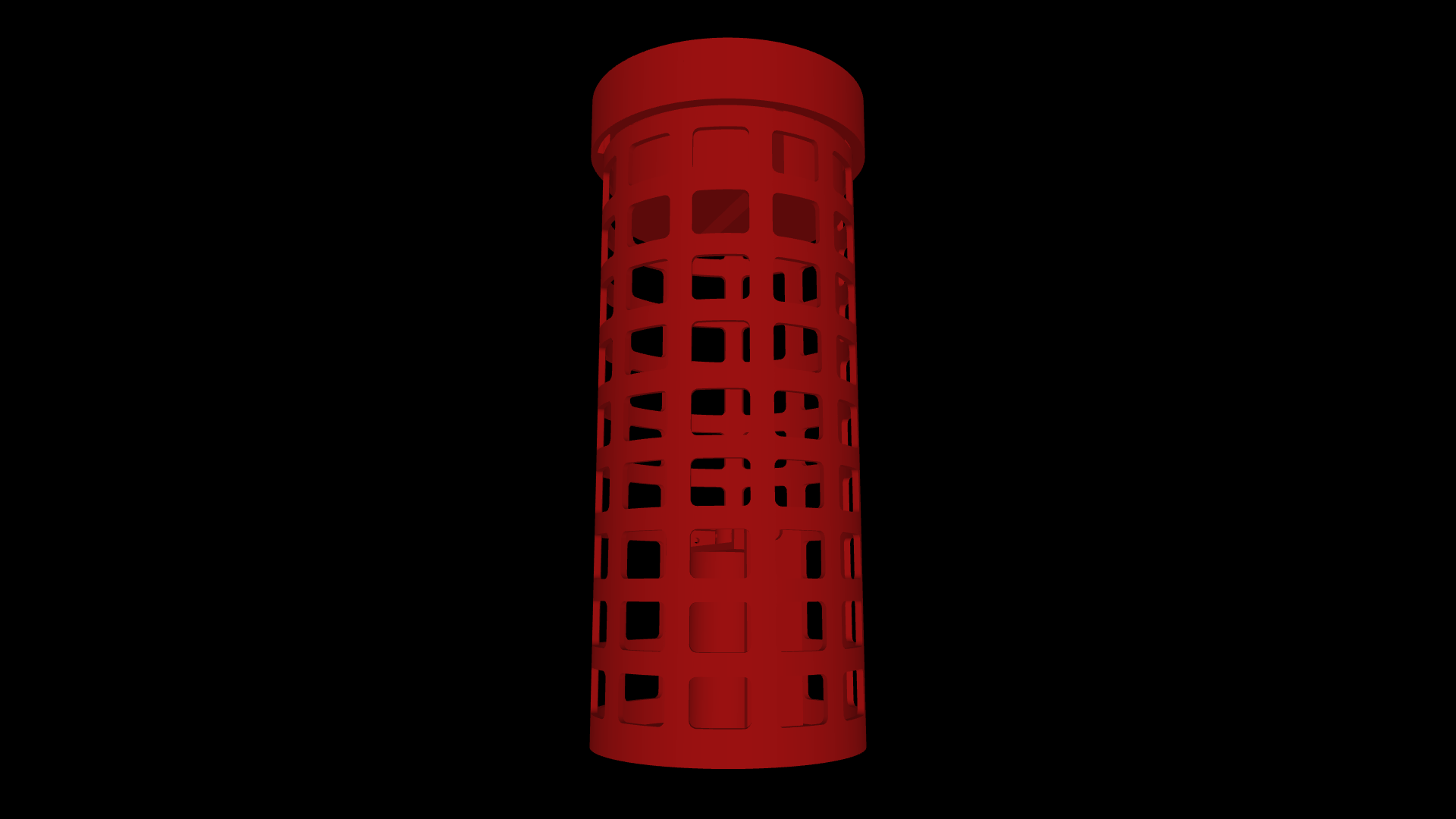}
    \caption{CANSAT CAD Model}
    \label{fig:enter-label}
\end{figure}

\subsection{Descent Control Subsystem(DCS)}
The Aim of DCS is to maintain a constant velocity and land the Cansat smoothly to the ground with the help of Parachute deployed from a Drone.Designing of a good parachute is mandatory. The Elements of a DCS is Parachute Material selection, Type of Parachute,way of holding the Parachute, the way we fold the Parachute and building a recovery system that prevents any casualties.
\subsubsection{Decent Rate Estimation}
\paragraph{When the parachute is deployed there are forces like drag and gravitational forces which show impact on the parachute. We use the drag equation given in Equation 1 for the design of the parachute. This formula is used for the understanding of the CANSAT decent.}

 \begin{equation}
D > \frac{1}{2} C_d \cdot P \cdot V^2 \cdot A
\end{equation}
\paragraph{Where P=density of air (1.225kg/m3), V=velocity (CANSAT), Cd=coefficient of drag, A= area of parachute, D = drag force. When the drag is equal to the weight of the CANSAT then the force becomes zero. Hence Weight equals equation 1. From Equation 2 we find the Velocity of the CANSAT that descends.}
\begin{equation}
V = \sqrt{\frac{2W}{C_d pA}} \quad 
\end{equation}
Nylon was the Parachute material chosen because of its balance of low density, sufficient tensile strength and low stiffness (low Young's modulus),  good elasticity property, and ability to protect against friction, indicating it can absorb shocks. Spill hole parachute was chosen because of its high Drag coefficient, High Durability when deployed, and also increases the stability by decreasing the drift. It was also maintained that the diameter of the Spill Hole is 20 percent of the Diameter of the Parachute, this will help in the stability of the CANSAT. The quality of the nylon also affects the stability of the CANSAT. The Arduino Nano, which is connected to the servo motors.When the BME180 reaches the 500m altitude then the parachutes are deployed from the CANSAT.The gyroscope works with help of the MPU6050 which is used to know the axis of the CANSAT and BO motors, which help stabilize the CANSAT by detumbling it.

\subsection{Electrical Power System (EPS)}
The Objective of EPS is to power the Cansat for its functioning during the flight. The function of each block of EPS is given in Table II. Lithium Ion Cell with a battery capacity of 1090 mAh Battery placed on a 5. 5mm x 2.1mm Adapter Port for UNO Power Jack. The nominal voltage of 12v is the output of the battery pack. A switch is mechanically operated to turn on the CANSAT. The battery pack should power the gyroscope,sensors and the parachute deployment system.
MQ135,MPU6050,BME180,servo motors,Neo 6M,Xbee S2c pro, these were connected to the 3.3v rail.The arduino Uno and Nano are connected with the 5v rail.The BO motors need 12v battery supply so it is directly connected to the battery.The Figure 3 shows the Block diagram of EPS.The CANSAT will be turned ON before putting it in the drone.Once all the subsystems are turned on the data keeps coming to the ground station which helps in giving real-time data and position of CANSAT.
The Table 2 explains all the functions.

\begin{table}
\centering

\begin{tabular}{|l|l|}
\hline
Component & Purpose \\
\hline
Battery & To supply power to the components \\
\hline
Switch & To connect and disconnect \\
&the battery from the circuit \\
\hline
Voltage Regulator & Adjusts the voltage \\
&as per the requirement \\
&(12V to 5V) \\
\hline
Microcontroller & Receives and \\
&processes data obtained from the sensor and \\
&to control the deployment of Parachutes \\
\hline
Sensors & Provides environment measurement \\
&and gives real-time \\
 & Data to the Microcontroller\\
\hline
Xbee & Sends and receives data from ground station \\
\hline
Parachute Deployment & Deployment of the parachutes  \\
& at the height of 500m using servo motors\\
\hline
Camera & To record the complete process \\
\hline
\end{tabular}
\caption{Functions of EPS}
\label{tab:my_table}
\end{table}

\subsection{On Board Computer Subsystem}
The On board computer is to process the data collected from the various sensors on the satellite and send them to the Xbee module for transmission. Arduino Uno acts as the on board computer in a NAMBI-VJ CANSAT. It uses ATMEGA328P.It has 14 digital input/output pins , 6 analog inputs, a 16 MHz ceramic resonator.The Data from the primary sensor is received by the OBC and sent it to On board Communication module called Xbee and sends it to ground station during flight. The On Board computer powers on and works on modes of operation. There are four modes of operation.
These Xbee modules are used in the transmitter and receiver which will 
send the data from the CANSAT to the Ground station.

\subsubsection{Modes of Operation}

\begin{itemize}
    \item \textbf{Mode 1}: CANSAT is launched to an altitude of 900 m by a drone.The CANSAT will be already turned ON in the start when it is loaded into the drone.  The CANSAT is in Descent towards the ground. Deployment of CANSAT is called Mode 1.
    \item \textbf{Mode 2}: A parachute is connected to an eye hook which is the primary parachute with descent rate of 10m/s to 12m/s.
    \item \textbf{Mode 3}: The secondary parachute is deployed using the servo motors which will flap out the parachutes at the altitude of 500 m which is known through the BME 180.The spin hole parachute with 15 cm radius is launched out of the can.The descent rate is 1-3m/s.
    \item \textbf{Mode 4}: Once all the parachutes are launched then the CANSAT goes through a lot of shock and instability.To overcome that, the gyroscope will be running from the height of 500 m which will help in increasing the stability of the CANSAT.This also helps in reducing the drift of the satellite and the supports soft landing.
    \item \textbf{Mode 5}: The sensed data by the sensor are converted into digital in case of analog and transmitted to the ground station. The ground station will receive the data for every 2 seconds.
    \item \textbf{Mode 6}: The data will be scraped from the serial monitor using Python code and will be pushed into the FIREBASE which is a cloud database. Once the data is sent from the CANSAT it will be a buzzer sound so that it verifies whether it is sending data or not.Once it hits the ground as per the altitude sensor then is continuously beeps. This will be helpful at the time of recovery of the CANSAT.Figure 4 show the graphs. As we receive the data packets from the CANSAT it will show longitude and latitude, PPM values, Temperature, Power consumption, Pressure, Altitude,(x,y,z) axis rotation, and acceleration of the CANSAT from the launch.

\end{itemize}
\begin{figure}
    \centering
    \includegraphics[width=1\linewidth]{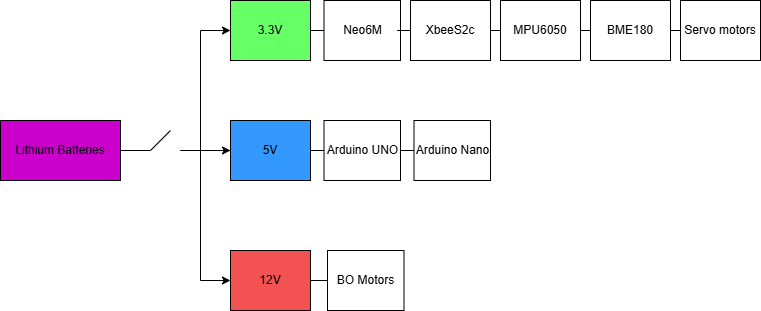}
    \caption{EPS Block Diagram}
    \label{fig:enter-label}
\end{figure}
\subsection{On Board Communication}
Xbee Module is used for Data communication with the satellite as Ground Station. XBee is a module produced by Digi International mainly used as  a radio communication transceiver and receiver. It is a mesh communication protocol that sits on top of IEEE 802.15.4 ZigBee standard. In XBee S2C Pro communication rate is  2.4GHz speed.This has a range of 1.2km with an omnidirectional antenna.Table 3 shows the configuration of Xbee S2C Pro with Arduino Uno which acts as a Transmitter.The Xbee operates at 3.3V and should be configured using XCTU Software.
\begin{table}
\centering

\begin{tabular}{| l | l |}
\hline
Xbee S2C Pro & Arduino Uno \\
\hline
3.3v & 3.3v \\
\hline
Gnd & Gnd \\
\hline
Tx & D2 \\
\hline
Rx & D3 \\
\hline

\end{tabular}
\caption{Pin Configuration with Uno}
\label{tab:my_table}
\end{table}

\subsection{Ground Station Subsystem}
The use of the ground station is to receive and transmit data from CANSAT during Flight. The block Diagram of Ground station is shown in figure 6. The Data from the CANSAT is transmitted via Xbee S2C pro.The Ground station has a Xbee module that is configured in Receive mode. Received data is sent to the processing unit of the ground station made of Arduino Nano.The data is first displayed on the serial monitor of the arduino IDE.Then this data is scrapped using the python code and then pushed in to the database.This database will be updated as per the data coming from the serial monitor.In the front end Next.js is used and in the back end Node.js.This data is visible in the vercel app.The Figure 4 shows the GUI of GS.
\begin{figure}
    \centering
    \includegraphics[width=0.75\linewidth]{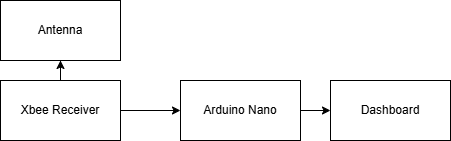}
    \caption{Ground Station Block Diagram}
    \label{fig:enter-label}
\end{figure}

\section{Experimental Results}
Before the launch test, we have performed a couple of hardware simulations. We have tested the transmitter and the receiver circuits by varying the distance between them from around 0m to 900m. The MQ135 sensor, BME180 sensor, GPS Neo 6m module,MPU6050, Xbee S2C Pro module and the Servo motor were connected to the Arduino Uno to form the transmitter circuit. The Arduino UNO was connected to a Xbee module to form the receiver circuit. The transmitted data and the received data were observed on the GUI. After the breadboard tests, we have performed a few drop tests from one of the buildings of our college. The testing were done to know which type of parachute has a good descent rate which can help in soft landing of the CANSAT. The descent of the parachute is 3m/s.

\begin{figure}
    \centering
    \includegraphics[width=1.\linewidth]{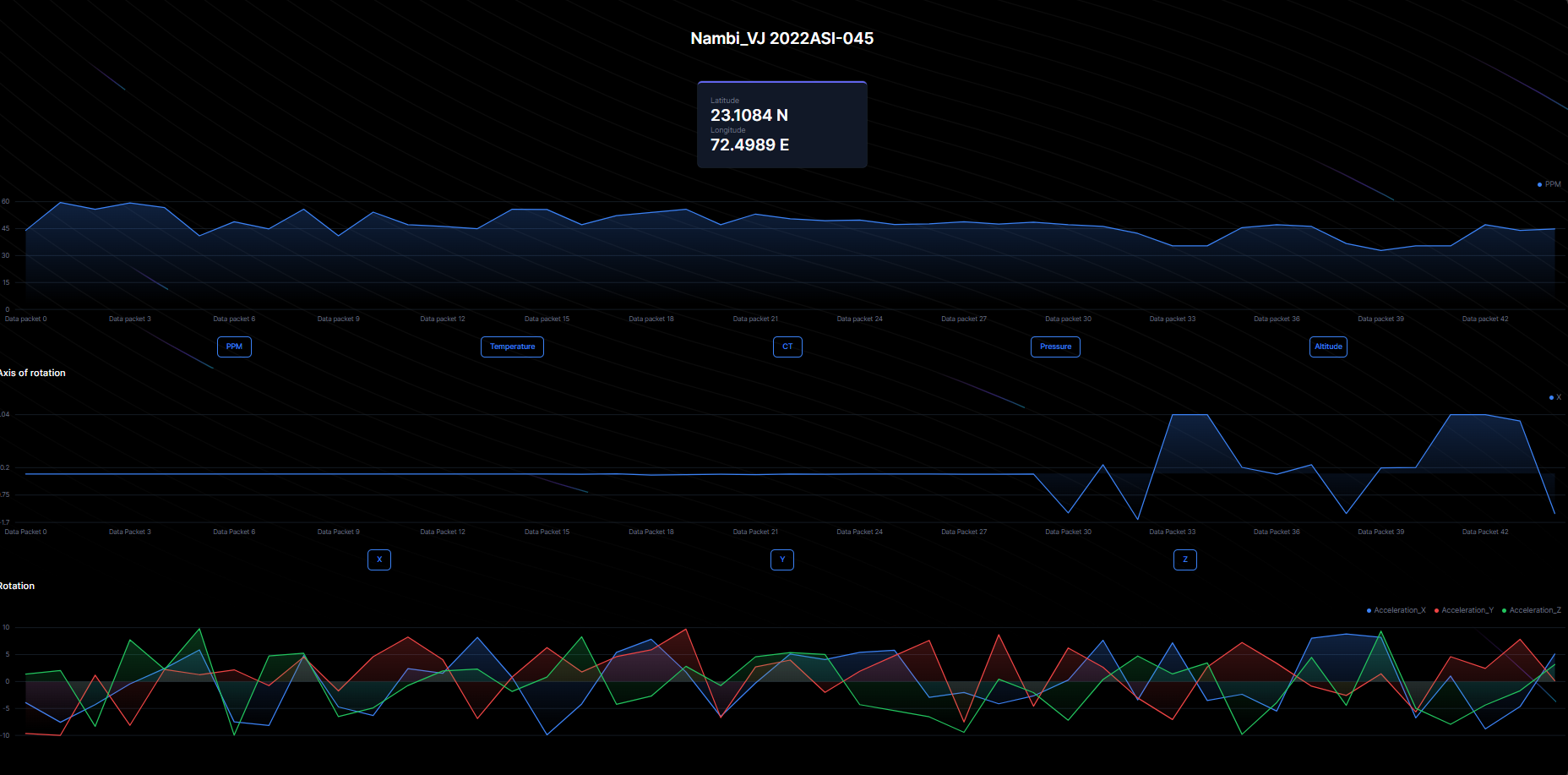}
    \caption{Ground Station GUI }
    \label{fig:enter-label}
\end{figure}
\subsection{Experiment at Ahmedabad, Gujarat}
On April 16th 2024,Drone has 5 cm greater than that of the CANSAT for easy insertion. Once the CANSAT was secure in the Drone, we launched the Satellite to a height of 900m. During the descent, the Parachutes of the CANSAT were deployed, the electronics of the CANSAT were turned on and servo motors opened the parachutes. Following this, the telemetry data from the CANSAT was collected at the Ground Station with the help of Xbee communication module. Simultaneously,The results of the experiment were observed at the GUI Dashboard. The descent of CANSAT was timed to be approximately 25 minutes. The Snapshot of experiments at Ahmedabad,Gujarat.The Table 4 shows the results after the experiments.
 \begin{figure}
    \centering
    \includegraphics[width=1\linewidth]{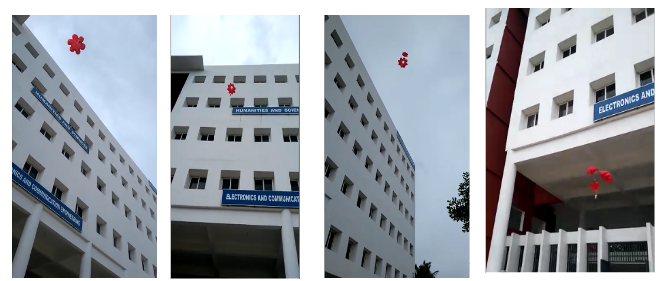}
    \caption{Parachute Testing }
    \label{fig:enter-label}
\end{figure}

\section{Conclusion}
Thus a Novel CANSAT was designed for 727gms tested and operated at an altitude of 900 m from the ground level and intended actions such as power on of the Satellite after deployment from the satellite, Antenna Deployment, Sensor information of Air quality Monitoring and Temperature with latitude and longitude of a decent path were tested, deployed and verified. 
 \begin{table}
\centering

\begin{tabular}{|l|l|l|l|l|l|l|l|}
\hline
Latitude & Longitude & Temp & Altitude & X & Y  & PPM \\
\hline
23.11 & 72.49 & 41.3 & 150.44 & -0.02 & -0.02 & 44 \\
23.11 & 72.49 & 41.4 & 158.36 & -0.02 & -0.02 & 59.54 \\
23.11 & 72.49 & 41.4 & 250.11 & -0.02 & -0.02 & 55.8 \\
23.11 & 72.49 & 41.4 & 269.86 & -0.02 & -0.02 & 40.98 \\
23.11 & 72.49 & 41.5 & 353.69 & -0.02 & -0.02 & 48.82 \\
23.11 & 72.49 & 41.5 & 359.61 & -0.02 & -0.02 & 55.8 \\
23.11 & 72.49 & 41.5 & 457.94 & -0.02 & -0.02 & 40.98 \\
23.11 & 72.49 & 41.5 & 469.03 & -0.02 & -0.02 & 47.17 \\
23.11 & 72.49 & 41.6 & 528.28 & -0.02 & -0.02 & 55.8 \\
23.11 & 72.49 & 41.6 & 569.78 & -0.03 & -0.02 & 47.17 \\
23.11 & 72.49 & 41.6 & 678.90 & -0.01 & -0.03 & 52.22 \\
23.11 & 72.49 & 41.7 & 690.17 & -0.06 & -0.02 & 53.99 \\
23.11 & 72.49 & 41.7 & 720.94 & -0.04 & -0.05 & 55.8 \\
23.11 & 72.49 & 41.7 & 770.38 & -0.03 & -0.01 & 47.17 \\
\hline
\end{tabular}
\caption{Results at the Gujarat Launch}
\label{tab:my_table}
\end{table}





 \bibliographystyle{IEEEtran}

\begin{thebibliography}{8}

\bibitem{8985514} 
Ramadhan, Rizki Pratama and Ramadhan, Aditya Rifky and Putri, Shindy Atila and Latukolan, Merlyn Inova Christie and Edwar and Kusmadi, 
"Prototype of CanSat with Auto-gyro Payload for Small Satellite Education," in \textit{2019 IEEE 13th International Conference on Telecommunication Systems, Services, and Applications (TSSA)}, pp. 243-248, 2019, doi: 10.1109/TSSA48701.2019.8985514.

\bibitem{10074552} 
Shukla, Prakhar and Mishra, Raj and Sardar, Uzair Ahmad and Mohapatra, B., 
"Satellite Design for CANSAT with Autorotatig payloads," in \textit{2022 4th International Conference on Advances in Computing, Communication Control and Networking (ICAC3N)}, pp. 2385-2392, 2022, doi: 10.1109/ICAC3N56670.2022.10074552.

\bibitem{6581344} 
Bulut, Sultan Nur and Gül, Mahircan and Beker, Can and İpek, İbrahim İlge and Koçulu, Ömer Eren Can and Topaloğlu, Çınar and Dinçer, Nurullah and Kırlı, Ahmet and Ertuğrul, Hasan Fatih and Tüfekci, Celal Sami, 
"Model satellite design for CanSat Competition," in \textit{2013 6th International Conference on Recent Advances in Space Technologies (RAST)}, pp. 913-917, 2013, doi: 10.1109/RAST.2013.6581344.

\bibitem{10212104} 
Kaur, Roopmeet and Dash, Biswajeet and Shiney O, Jeba and Singh, Sukhpreet, 
"Revolutionizing CanSat Technology with Vega Processors: A Comparative Study," in \textit{2023 2nd International Conference on Edge Computing and Applications (ICECAA)}, pp. 1276-1282, 2023, doi: 10.1109/ICECAA58104.2023.10212104.

\bibitem{8399591} 
Ostaszewski, Michał and Dzierzek, Kazimierz and Magnuszewski, Łukasz, 
"Analysis of data collected while CanSat mission," in \textit{2018 19th International Carpathian Control Conference (ICCC)}, pp. 1-4, 2018, doi: 10.1109/CarpathianCC.2018.8399591.

\bibitem{10528162} 
Reddy, Ramya and Panangipalli, Vishwanath Kumar and Mala, Narayana and Bairu, Charan Sunny and Rumale, Rujul, 
"AathreyaSat: A CanSat Model for Air Pollution Measurement in Competition," in \textit{2024 IEEE Wireless Antenna and Microwave Symposium (WAMS)}, pp. 1-5, 2024, doi: 10.1109/WAMS59642.2024.10528162.

\bibitem{8994718} 
Islam, Tooba and Noureen, Ayesha and Mughal, Muhammad Rizwan and Nadeem, M. Asim, 
"Design and Development of a Weather Monitoring Satellite, CanSat," in \textit{2019 15th International Conference on Emerging Technologies (ICET)}, pp. 1-6, 2019, doi: 10.1109/ICET48972.2019.8994718.

\bibitem{8002947} 
Kizilkaya, Muhterem Özgür and Oğuz, Abdullah Ersan and Soyer, Süleyman, 
"CanSat descent control system design and implementation," in \textit{2017 8th International Conference on Recent Advances in Space Technologies (RAST)}, pp. 241-245, 2017, doi: 10.1109/RAST.2017.8002947.
\bibitem{10245113} 
Sinha, Madhurima and R, Lakshya and R, Pranitha and K, Srikanth and Raj, Kunal and K, Vasanth, 
"Design of Trash Can Sized Satellite for Air Quality Monitoring at an Altitude of 300m Above Ground Level," in \textit{2023 International Conference on Circuit Power and Computing Technologies (ICCPCT)}, pp. 88-95, 2023, doi: 10.1109/ICCPCT58313.2023.10245113.

\bibitem{7362931} 
Bautista-Linares, Efren and Morales-Gonzales, Enrique A. and Herrera-Cortez, Mario and Narvaez-Martinez, Esther A. and Martinez-Castillo, Jaime, 
"Design of an Advanced Telemetry Mission Using CanSat," in \textit{2015 International Conference on Computing Systems and Telematics (ICCSAT)}, pp. 1-4, 2015, doi: 10.1109/ICCSAT.2015.7362931.

\bibitem{10441654} 
Sharma, Harsh and Sehgal, Abhinav and Jindal, Harsh and Dutta, Aditi and Sharma, Bhawna, 
"Designing and Developing a CanSat for Environmental Monitoring and Scientific Exploration," in \textit{2023 International Conference on Advanced Computing \& Communication Technologies (ICACCTech)}, pp. 358-363, 2023, doi: 10.1109/ICACCTech61146.2023.00065.

\bibitem{10436175} 
Rivadeneira, Franco and Godinez, Diego and Kiyan, Kioshi and Huayapa, Victor and Acosta, Sebastian and Perez, Nicole and Hinostroza, Abel and Arce, Diego, 
"Run2Sat I: Design and Implementation of a CanSat with Autogyro System," in \textit{2023 IEEE Colombian Caribbean Conference (C3)}, pp. 1-6, 2023, doi: 10.1109/C358072.2023.10436175.

\bibitem{9231031} 
Hasan Raian, F.M. Tanvir and Islam, H.M. Jahirul and Islam, Md. Saiful and Azam, Rafiul and Islam, H.M. Jahidul and Debnath, Sutapa, 
"An Affordable CanSat Design and Implementation to Study Space Science for Bangladeshi Students," in \textit{2020 IEEE Region 10 Symposium (TENSYMP)}, pp. 1205-1208, 2020, doi: 10.1109/TENSYMP50017.2020.9231031.

\bibitem{10398763} 
Chun, Carrington and Kihei, Billy and Chakravarty, Sumit and Tanveer, M. Hassan, 
"Open-Source, Low-Cost, and Bimodal Amateur Radio Communication Paradigm for Scientific CanSats," in \textit{2023 6th International Conference on Robotics, Control and Automation Engineering (RCAE)}, pp. 212-218, 2023, doi: 10.1109/RCAE59706.2023.10398763.

\bibitem{10374870} 
Chun, Carrington and Patel, Uday and Tanveer, M. Hassan and Swift, Tom and Dallesasse, Kevin and Chakravarty, Sumit, 
"Crafting CanSats: A Novel Modular Design Paradigm for Scientific CanSats," in \textit{2023 IEEE 20th International Conference on Smart Communities: Improving Quality of Life using AI, Robotics and IoT (HONET)}, pp. 68-72, 2023, doi: 10.1109/HONET59747.2023.10374870.

\end{thebibliography}

\end{document}